\documentclass[aps,showpacs,showkeys]{revtex4}
\usepackage{amsmath,amsthm,amssymb,epsfig,alltt}

\begin{document}

\def\a{\alpha}
\def\b{\beta}
\def\d{{\delta}}
\def\l{\lambda}
\def\e{\epsilon}
\def\p{\partial}
\def\m{\mu}
\def\n{\nu}
\def\t{\tau}
\def\th{\theta}
\def\s{\sigma}
\def\g{\gamma}
\def\G{\Gamma}
\def\o{\omega}
\def\r{\rho}
\def\D{\Delta}
\def\half{\frac{1}{2}}
\def\hatt{{\hat t}}
\def\hatx{{\hat x}}
\def\hatp{{\hat p}}
\def\hatX{{\hat X}}
\def\hatY{{\hat Y}}
\def\hatP{{\hat P}}
\def\haty{{\hat y}}
\def\whatX{{\widehat{X}}}
\def\whata{{\widehat{\alpha}}}
\def\whatb{{\widehat{\beta}}}
\def\whatV{{\widehat{V}}}
\def\hatth{{\hat \theta}}
\def\hatta{{\hat \tau}}
\def\hatrh{{\hat \rho}}
\def\hatva{{\hat \varphi}}
\def\barx{{\bar x}}
\def\bary{{\bar y}}
\def\barz{{\bar z}}
\def\baro{{\bar \omega}}
\def\barpsi{{\bar \psi}}
\def\sp{\sigma^\prime}
\def\nn{\nonumber}
\def\cb{{\cal B}}
\def\2pap{2\pi\alpha^\prime}
\def\wideA{\widehat{A}}
\def\wideF{\widehat{F}}
\def\beq{\begin{eqnarray}}
 \def\eeq{\end{eqnarray}}
 \def\4pap{4\pi\a^\prime}
 \def\xp{{x^\prime}}
 \def\sp{{\s^\prime}}
 \def\ap{{\a^\prime}}
 \def\tp{{\t^\prime}}
 \def\zp{{z^\prime}}
 \def\xpp{x^{\prime\prime}}
 \def\xppp{x^{\prime\prime\prime}}
 \def\barxp{{\bar x}^\prime}
 \def\barxpp{{\bar x}^{\prime\prime}}
 \def\barxppp{{\bar x}^{\prime\prime\prime}}
 \def\barchi{{\bar \chi}}
 \def\baro{{\bar \omega}}
 \def\bpsi{{\bar \psi}}
 \def\barg{{\bar g}}
 \def\barz{{\bar z}}
 \def\bareta{{\bar \eta}}
 \def\ta{{\tilde \a}}
 \def\tb{{\tilde \b}}
 \def\tc{{\tilde c}}
 \def\tz{{\tilde z}}
 \def\tJ{{\tilde J}}
 \def\tpsi{\tilde{\psi}}
 \def\tal{{\tilde \alpha}}
 \def\tbe{{\tilde \beta}}
 \def\tga{{\tilde \gamma}}
 \def\tchi{{\tilde{\chi}}}
 \def\barth{{\bar \theta}}
 \def\bareta{{\bar \eta}}
 \def\barom{{\bar \omega}}
 \def\bole{{\boldsymbol \epsilon}}
 \def\bolth{{\boldsymbol \theta}}
 \def\bomega{{\boldsymbol \omega}}
 \def\bolmu{{\boldsymbol \mu}}
 \def\bola{{\boldsymbol \alpha}}
 \def\bolb{{\boldsymbol \beta}}
 \def\bolX{{\boldsymbol X}}
 \def\mathN{{\boldsymbol n}}
 \def\bba{{\boldsymbol a}}
 \def\bbA{{\boldsymbol A}}
 \def\mathP{{\mathbb P}}
 \def\mathN{{\boldsymbol N}}
 \def\mathN{{\mathbb N}}
 \def\bbP{{\boldsymbol P}}

\setcounter{page}{1}
\title[]{Deformation of the Cubic Open String Field Theory
}

\author{Taejin Lee}
\affiliation{
Department of Physics, Kangwon National University, Chuncheon 200-701
Korea}

\email{taejin@kangwon.ac.kr}

\begin{abstract}
We study a consistent deformation of the cubic open bosonic string theory in such a way 
that the non-planar world sheet diagrams of the perturbative string theory are  
mapped onto their equivalent planar diagrams of the light-cone string field theory 
with some length parameters fixed.
An explicit evaluation of the cubic string vertex in the zero-slope limit yields the correct 
relationship between the string coupling constant and the Yang-Mills coupling constant. 
The deformed cubic open string field theory is shown to produce the non-Abelian Yang-Mills action 
in the zero-slope limit if it is defined on multiple D-branes. Applying the  
consistent deformation systematically to multi-string world sheet diagrams, we may be 
able to calculate scattering amplitudes with an arbitrary number of external open strings.

\end{abstract}




\maketitle

\section{Introduction}

If its perturbation theory is 
correctly defined, the covariant string field theory is expected to replace eventually the quantum field theory which has not been successful to describe quantum particles with spin two and higher spins. However, in practice, it is rather difficult to make use of the covariant cubic string field theory \cite{Witten1986,Witten92p} to calculate the particle scattering amplitudes. The main reason is that the world sheet diagrams of cubic open string field theory are 
non-planar unlike those of the light-cone string field theory \cite{Mandelstam1973, Mandelstam1974, Kaku1974a, Kaku1974b, Cremmer74, Cremmer75, GreenSW}. Witten \cite{Witten1986} introduced an associative product between the open string field operators which represents the mid-point overlapping interaction.
With the associative star product, the string field action takes the form of the Chern-Simons three-form
which is invariant under the BRST gauge transformation. The cubic open string field theory has a merit 
of the BRST gauge invariance due to the associative algebra of the string field operators. But at the 
same time the mid-point overlapping interaction renders the world-sheet diagrams non-planar so that 
it becomes a difficult task to get the Fock space representations of the multi-string vertices.

The Fock space representation of the three-string vertex of the cubic open string field theory has been obtained by Gross and Jevicki in Refs. \cite{Grossjevicki87a} and \cite{Grossjevicki87b} by mapping the world-sheet diagram of six strings onto a circular disk
and imposing an orbifold condition. The conformal mapping of the four-string world sheet to the upper half complex plane with branch cuts has been constructed by Giddings \cite{Giddings86}. The Neumann functions of the three-string vertex have been calculated in Refs. \cite{Cremmer86,Grossjevicki87a,Samuel86} and the Neumann functions of the four-string vertex has been computed by Samuel in Ref. \cite{Samuel88}.
However, there seems to be no similarity between the conformal mappings for the three-string vertex 
and that of the four-string vertex. It seems also difficult to apply those constructions 
of the conformal mappings to more complex world sheet diagrams of multi-string vertices. 
Thus, it is desirable to 
develop a more systematic technique which could be applied to string scattering diagrams with an arbitrary 
number of external strings. In the present work, we propose a consistent deformation of the world sheet 
diagrams which transforms the non-planar diagrams of multi-string scattering 
into planar diagrams. Once having obtained the planar diagrams of the multi-string vertices, we can make use of the light-cone string field theory technique by mapping the world sheet diagrams onto the upper half complex plane. For the three-string vertex and the four-string vertex, it is enough to choose external string states such that physical string states are encoded only on the halves of the external strings. By an explicit calculation, we shall show that 
the deformed cubic string vertex yields the three-gauge field vertex with the correct Yang-Mills coupling 
constant in the zero-slope limit. The four-gauge field vertex of the Yang-Mills action shall be also evaluated by using the deformed world sheet diagram of the four-string vertex which is effectively generated by two cubic 
string vertices and an intermediate string propagator.  

\section{Deformation of the Witten's Open String Field Theory Diagrams}

We shall begin the Witten's cubic open string field theory action \cite{Witten1986} on muti-D-branes which is given as 
\beq
{\cal S} = \int \text{tr} \left( \Psi * Q \Psi + \frac{2g}{3} \Psi * \Psi * \Psi \right)
\eeq
where $Q$ is the BRST operator and the sring field $\Psi$ is $U(N)$ matrix valued 
\beq
\Psi = \Psi^0 + \sum_a \Psi^a T^a, ~~~ a =1, \cdots, N^2-1 . 
\eeq
The star product of between the string field operators is defined as follows 
\beq
\left(\Psi_1 * \Psi_2\right) [X^{(3)}(\s)] &=& \int \prod_{\frac{\pi}{2} \le \s \pi} DX^{(1)}(\s) \prod_{0 \le \s \frac{\pi}{2}} DX^{(2)}(\s)  \nn\\
&& 
\prod_{\frac{\pi}{2} \le \s \le \pi} \d \left[X^{(1)}(\s) - X^{(2)}(\pi -\s) \right] \Psi_1[X^{(1)}(\s)] \Psi_2[X^{(2)}(\s)], \nn\\
X^{(3)}(\s) &=& \left\{ \begin{array}{l l l} 
X^{(1)}(\s)~~ & \text{for} & 0 \le \s \le \frac{\pi}{2}, \\
X^{(2)}(\s) ~~ & \text{for} &  \frac{\pi}{2} \le \s \le \pi. 
\end{array} \right. \label{star}
\eeq
In terms of the normal modes, the string coordinates $X^{(r)}(\s)$, $r=1,2,3 $ are expanded as 
\beq \label{mode}
X^{(r)}(\s) 
&=& x^{(r)} + 2\sum_{n=1} \frac{1}{\sqrt{n}} x^{(r)}_n  \cos \left(n \s \right).
\eeq
It is the associativity of the star product algebra 
\beq
\left(\Psi_1 * \Psi_2\right) * \Psi_3 = \Psi_1 * \left(\Psi_2 * \Psi_3 \right)
\eeq
that ensures invariance of the cubic string field theory action under the gauge transformation of the 
string field
\beq
\d \Psi = Q * \e + \Psi * \e - \e * \Psi .  
\eeq 

\begin{figure}[htbp]
   \begin {center}
    \epsfxsize=0.3\hsize

	\epsfbox{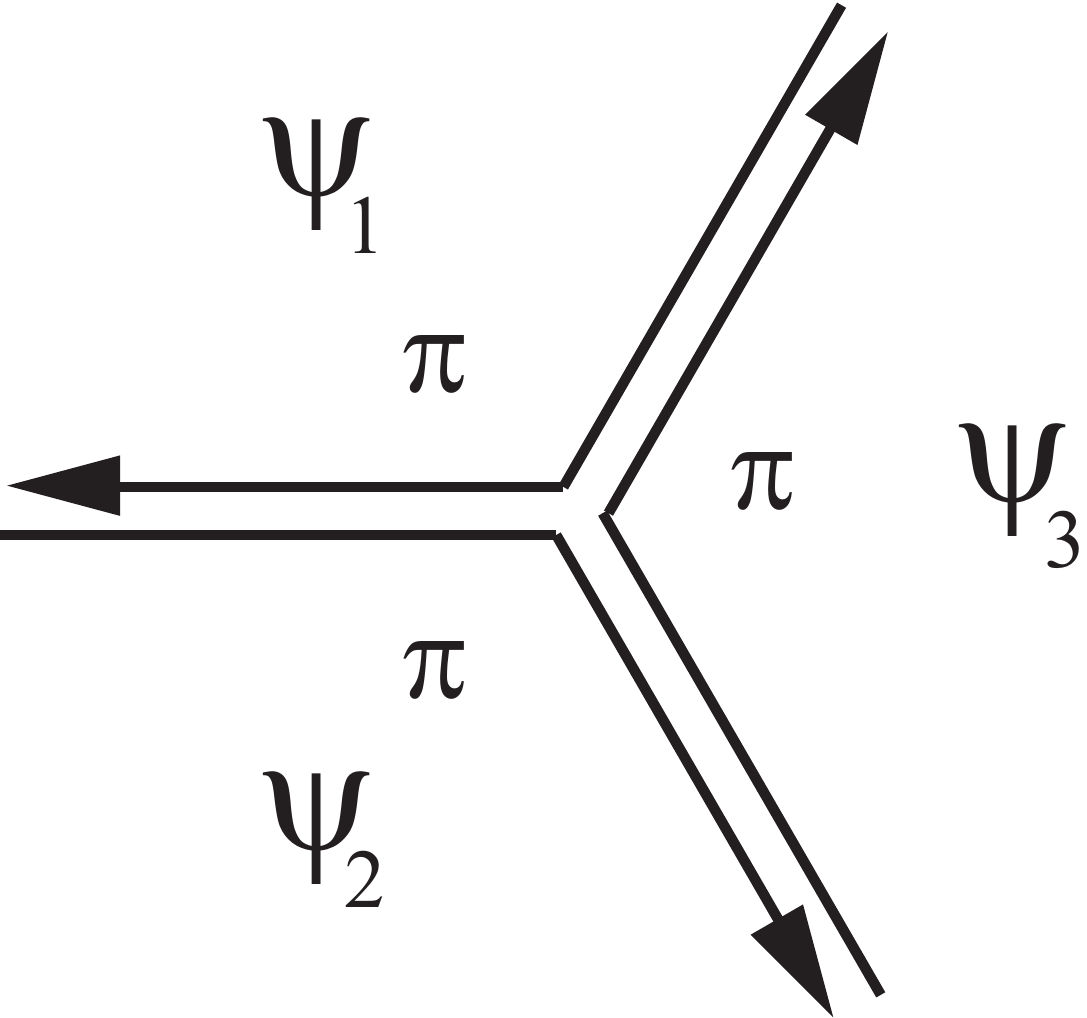}
   \end {center}
   \caption {\label{cubic} The Mid-point overlapping interaction of the cubic open string field theory}
\end{figure}

In order to discuss the deformation of the cubic string field theory we extend the 
range of the world sheet coordinate $\s$ firstly as  
\beq
0 \le \s \le \pi ~~\Longrightarrow ~~0 \le \s \le 2\pi
\eeq 
The mid-point is now located at $\s = \pi$. Accordingly, the star product Eq. (\ref{star}) and the 
normal mode expansion Eq. (\ref{mode}) should be appropriately redefined 
\beq
X^{(r)}(\s) 
&=& x^{(r)} + 2\sum_{n=1} \frac{1}{\sqrt{n}} x^{(r)}_n  \cos \left(\frac{n}{2} \s \right), ~~~
r = 1, 2, 3 . 
\eeq
as shown in Fig. \ref{cubic}. 

\begin{figure}[htbp]
   \begin {center}
    \epsfxsize=0.5\hsize

	\epsfbox{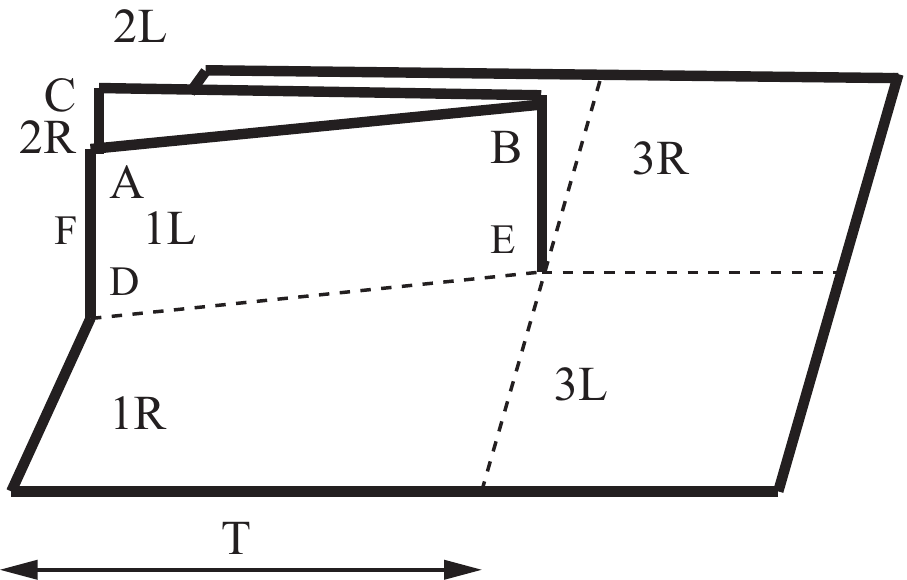}
   \end {center}
   \caption {\label{star} The world sheet diagram of the three-string scattering}
\end{figure}

\begin{figure}[htbp]
   \begin {center}
    \epsfxsize=0.5\hsize

	\epsfbox{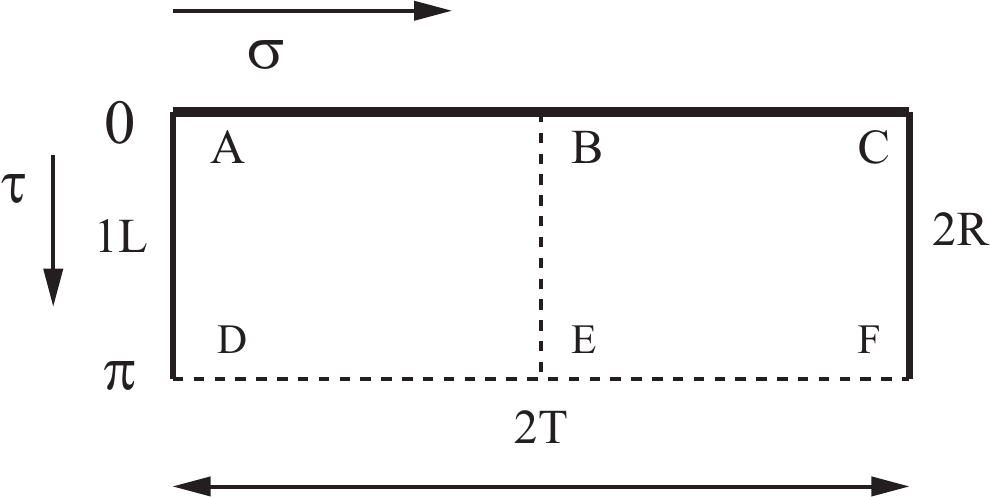}
   \end {center}
   \caption {\label{edge} Three-String Vertex Diagram of Witten's Cubic Open String Field Theory}
\end{figure}

Fig. \ref{star} depicts the world sheet diagram of three-string scattering.
We observe that during the scattering process, physical information encoded on the left half of the first string and 
physical information encoded on the right half of the second string are not carried over to the third string. 
In view of scattering process roles of the left half of the first string and the right half of the second 
string are auxiliary. 
Note that the strings satisfy the Neumann boundary condition on the boundary
$\overline{ABC}$ in Fig. \ref{cubic} .  We may separate the path, corresponding to the
world sheet trajectory of the left half of the first string and the right half of the second 
string from the rest part of the world sheet of three-string scattering. On the patch as we redefine the 
world sheet local coordinates by interchanging $\t \leftrightarrow \s$, the boundary condition on $\overline{ABC}$ becomes
\beq
\p_\s X^\m = 0  \rightarrow \p_\t X^\m = 0 . 
\eeq
(See fig. \ref{edge}.) 
On the patch we also define new string coordinates 
\beq
X^\m (\s) &=& x^\m + 2 \sum_{n=1} \frac{1}{\sqrt{n}} x^\m_n \cos \left(\frac{n\pi \s}{2T} \right) \nn\\
&=& x^\m + \sum_{n=1} \frac{i}{\sqrt{n}}\left(a_n^\m - a^{\m\dagger}_n \right) \cos \left(\frac{n\pi \s}{2T} \right). 
\eeq 

The Neumann condition on the boundary $\overline{ABC}$
may be written as
\beq
\p_\t X^\m \vert N \rangle &=& 0, \\
\vert N \rangle = c_N \exp\left(-\half \sum_{n=1} a^\dag_n \cdot a^\dag_n  \right) \vert 0 \rangle. 
\eeq
where $c_N$ is a normalization constant for the Neumann state.
$\vert N \rangle$ is the open string analogue of the Neumann boundary state of the closed string theory.  
The open string on the boundary $\overline{ABC}$ may propagates freely to the line $\overline{DEF}$ 
if the endpoints of the open string on the patch satisfy the Neumann condition 
\beq
\p_\s X^\m (0) = 0 ~~ \text{on}~ \overline{AD}~~~ \text{and}~ \p_\s X^\m (\pi) = 0, ~~~ \text{on}~ \overline{CF} .
\eeq
Then the open string state on $\overline{DEF}$ turns out to be the Neumann state again
\beq
\exp\left[-i \pi L_0 \right] \vert N \rangle =  - \vert N \rangle .
\eeq 
The extra phase factor $(-1)$ may be absorbed into the normalization constant of the Neumann state.
(We may also extend the range of $\s$ as $0 \le \s \le 2\pi$: It result in removing the factor $(-1)$ because
the open string state on $\overline{DEF}$ becomes $\exp\left[-i2\pi L_0 \right] \vert N \rangle = \vert N \rangle$.)
Hence, if we choose the Neumann condition for the left half of the first string and for the 
right half of the second string at the initial time, we may remove the 
patch which consists of the world sheets of the left half of the first string and the 
right half of the second string. The string path integral over the patch 
to scattering amplitude is simply 
\beq
\left\vert \langle N \vert e^{-i \pi (L_0 -1) } \vert N \rangle \right\vert =   1 .
\eeq
Thus, the string path integral over the patch does not contribute to the scattering amplitude

\begin{figure}[htbp]
   \begin {center}
    \epsfxsize=0.6\hsize

	\epsfbox{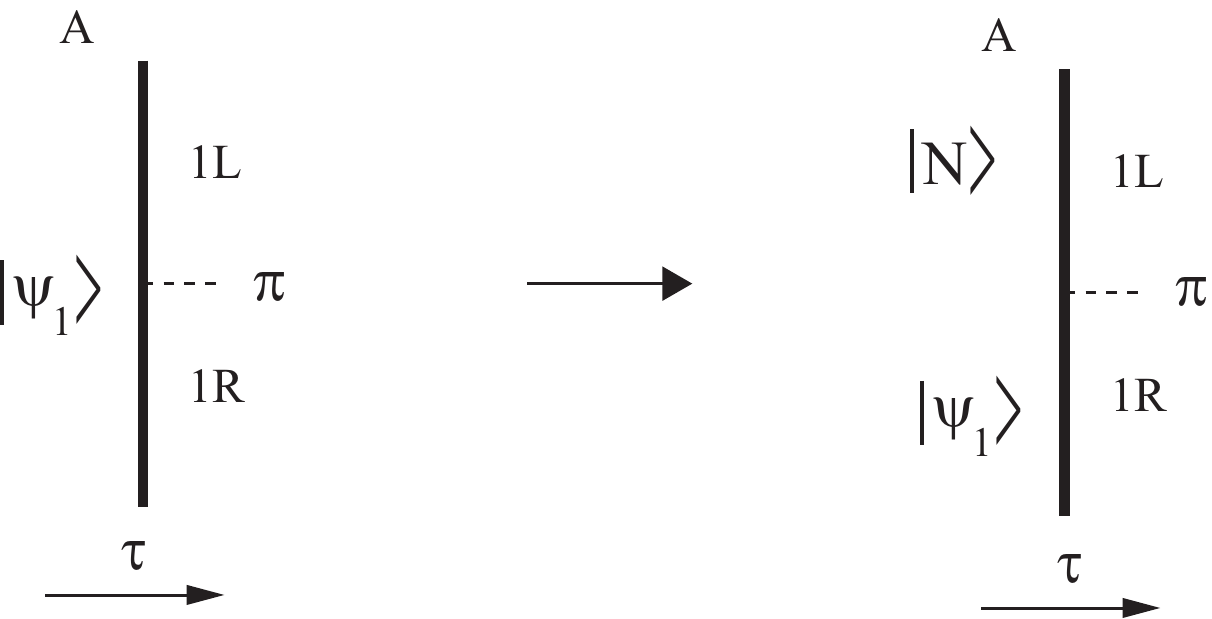}
   \end {center}
   \caption {\label{initial1} Deformation of the initial state of the first string}
\end{figure}

\begin{figure}[htbp]
   \begin {center}
    \epsfxsize=0.6\hsize

	\epsfbox{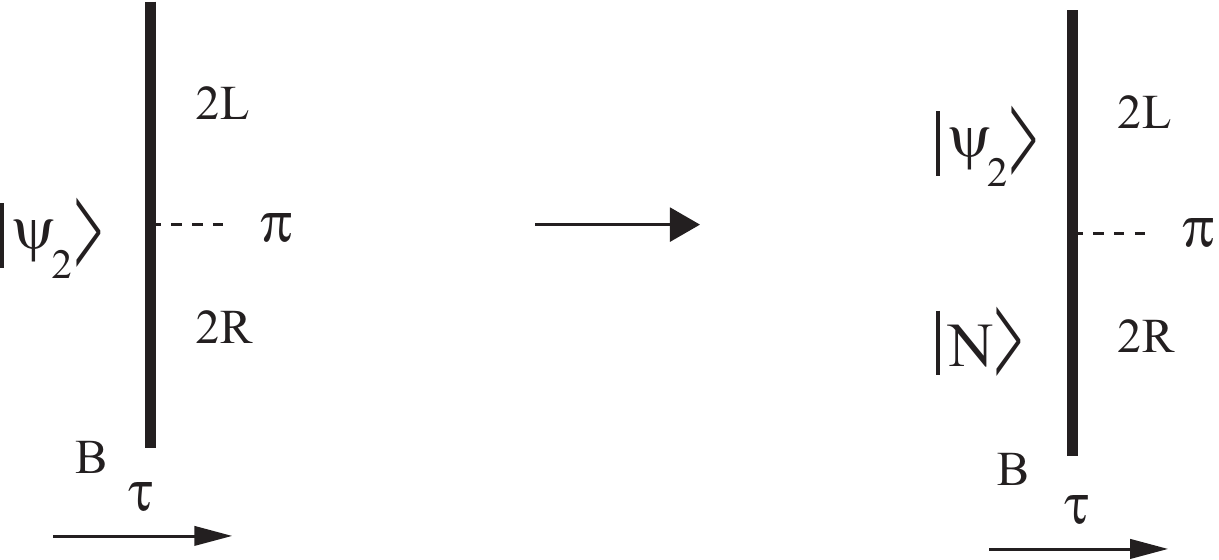}
   \end {center}
   \caption {\label{initial2} Deformation of the initial state of the second string}
\end{figure}

\begin{figure}[htbp]
   \begin {center}
    \epsfxsize=0.5\hsize

	\epsfbox{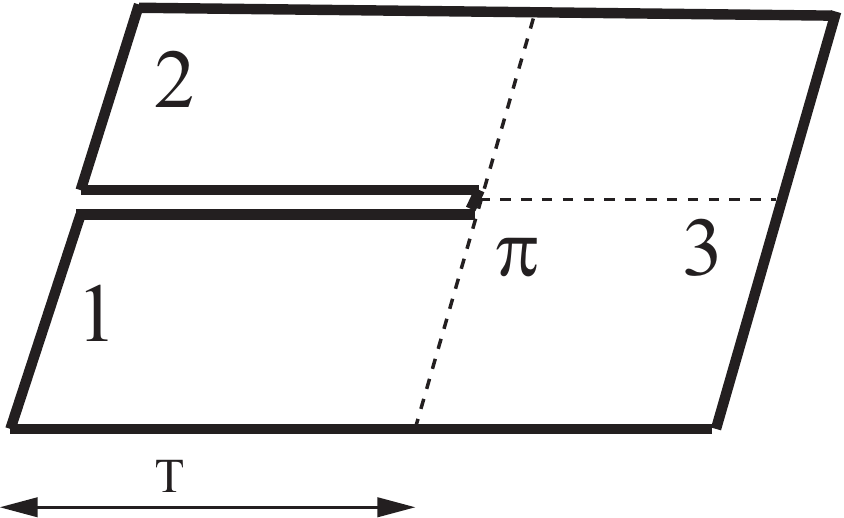}
   \end {center}
   \caption {\label{proper} Deformed world sheet diagram of the three-string scattering}
\end{figure}

To be consistent with this scheme we may encode the initial states of the 
first and the second string states onto the right half of the first string and the left
half of the second string respectively as depicted in Fig. \ref{initial1} and Fig. \ref{initial2}: 
\beq
\vert \Psi_1 \rangle  \rightarrow \vert N \rangle \otimes \vert \Psi_1 \rangle, ~~~
\vert \Psi_2 \rangle  \rightarrow \vert \Psi_2 \rangle \otimes \vert N \rangle 
\eeq 
Fig. \ref{proper} depicts the deformed world sheet diagram of the three-string scattering after
the auxiliary patch is completely removed. 
Because the world sheet diagram is not deformed uniformly, the associativity of the star product 
is not preserved. Consequently, the BRST gauge invariance is not manifest in the string field action with the
deformed cubic interaction. But if we formally keep the auxiliary patch, the associativity of the 
star product, hence the gauge invariance can be kept intact.  
As we remove the auxiliary patch, the world sheet diagram of the three-string becomes planar, which 
then can be mapped onto the upper half complex plane without any additional condition.

\section{Three-Gauge Field Vertex from the Deformed Three-String Vertex}

\begin{figure}[htbp]
   \begin {center}
    \epsfxsize=0.8\hsize

	\epsfbox{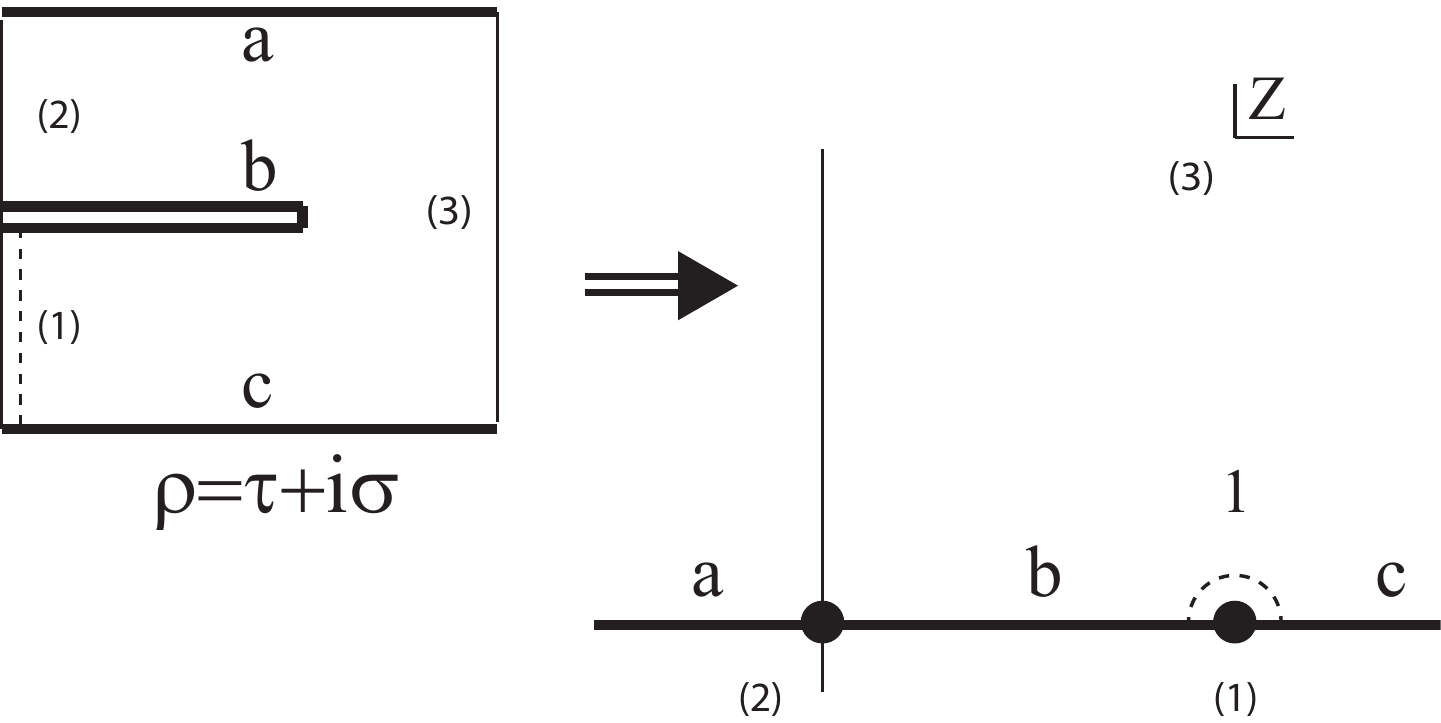}
   \end {center}
   \caption {\label{neumann3} Three-String scattering diagram is mapped onto the 
   upper half complex plane.}
\end{figure}

The planar diagram of the deformed three-string scattering is equivalent to that of the 
covariantized light-cone string field theory of HIKKO \cite{Hata1986} with length parameters fixed as 
\beq
\a_1 =1, ~~ \a_2 =1, ~~~ \a_3 =-2 .
\eeq 
Unlike the HIKKO's open covariant string field theory, we do not need to integrate over the
unphysical length parameters to make the string field action invariant under the BRST gauge transformation. 
Simply reattaching the auxiliary patch would restore the BRST gauge invariant form. 
On the planar world sheet we may introduce a global coordinate $\rho$ of 
which real part is the proper time $\text{Re} \rho = \t$. The planar world sheet may be mapped onto the
upper half complex plane by the Schwarz-Christoffel transformation given as  
\beq
\rho = \ln (z-1) + \ln z.
\eeq 
The temporal boundaries of the world sheet (labeled as $a, b, c$ in Fig. \ref{neumann3}) are mapped 
onto the real ine. On individual string world sheet patches we may define local coordinates $\zeta_r$, $r = 1, 2, 3$
which are related to $z$ as follows 
\begin{subequations}
\beq
e^{-\zeta_1} &=& e^{\t_0} \frac{1}{z(z-1)}, \\
e^{-\zeta_2} &=& - e^{\t_0} \frac{1}{z(z-1)} \\
e^{-\zeta_3} &=& - e^{- \frac{\t_0}{2}} \sqrt{z(z-1)} 
\eeq
\end{subequations} 
where $\t_0 = -2 \ln 2$. 

The Fock space representation of the three-string vertex in terms of the Neumann funcitons $\bar N^{rs}_{nm}$ follows from the light-cone string theory 
with length parameters fixed: 
\beq
\vert E_{[3]}[1,2,3] \vert 0 \rangle &=&  \exp \Biggl\{ \frac{1}{2} \sum_{r,s} \sum_{n, m \ge 1} \bar N^{rs}_{nm} \, 
\a^{(r)\dagger}_n \cdot \a^{(s)\dagger}_n + \sum_r \sum_{n \ge 1} \bar N^r_n \a^{(r)\dag}_n \cdot \bbP + \t_0 \sum_r \frac{1}{\a_r} \left(\frac{(p^{(r)})^2}{2} -1 \right) \Biggr\} \vert 0 \rangle , \nn\\
\bbP &=& p^{(2)} - p^{(1)} . 
\eeq 
The interaction term of three-string field may be written as 
\beq
{\cal S}_{[3]} &=& \frac{2g}{3} \langle \Psi_1, \Psi_2, \Psi_3 \vert E_{[3]}[1, 2, 3] \rangle. 
\eeq 
The three-gauge interaction term may be obtained by choosing the external state as 
\beq
\langle \Psi_1, \Psi_2, \Psi_3 \vert = \Bigl\langle 0 \Bigl \vert
\left\{\prod_{i=1}^3 A(p^{(i)}) \cdot a^{(i)}_1 \right\}
\eeq 
in the zero-slope limit:
\beq
S_{\text{Gauge}[3]} &=& \frac{2g_{YM}}{3} e^{- \t_0 \sum_{r=1}^3 \frac{1}{\a_r} }\int \prod_{i=1}^3 \frac{dp^{(i)}}{(2\pi)^d} (2\pi)^d \d \left(\sum_{i=1}^3 p^{(i)} \right) \nn\\
&& \text{tr} \Bigl\langle 0 \Bigl \vert
\left\{\prod_{i=1}^3 A(p^{(i)}) \cdot a^{(i)}_1 \right\} \half \sum_{r, s =1}^3\bar N^{rs}_{11} \left(a^{(r)\dag}_1 \cdot a^{(s)\dag}_1 \right) \sum_{t=1}^3 \bar N^{t}_1 \left(a^{(t)\dag}_1 \cdot 
\bbP \right)\Bigl \vert 0 \Bigl \rangle
\eeq 
where the Yang-Mills coupling constant $g_{YM}$ is related to the string interaction coupling $g$ as 
\beq
g_{YM} =\left(\ap\right)^{\frac{d}{4}-1} g. 
\eeq 
Making use of the explicit expressions of the Neumann functions
\begin{subequations}
\beq
\bar N^{11}_{11} &=& \frac{1}{2^4}, ~~~ \bar N^{22}_{11} = \frac{1}{2^4}, ~~~ \bar N^{33}_{11} = 2^2, \\
\bar N^{12}_{11} &=& \bar N^{21}_{11} = \frac{1}{2^4}, ~~~ \bar N^{23}_{11} = \bar N^{32}_{11} = \half, 
~~~ \bar N^{31}_{11} = \bar N^{13}_{11} = \half ,\\
\bar N^1_1 &=& \bar N^2_1 = \frac{1}{4}, ~~~ \bar N^3_1 = -1 ,
\eeq
\end{subequations}
we find the three-gauge interaction term
\beq
S_{\text{Gauge}[3]} &=& g_{YM}  \int \prod_{i=1} dp^{(i)} \d \left(\sum_{i=1}^3 p^{(i)} \right) p^\mu_1 
\,\text{tr} \Bigl( A^\n(p^{(1)})  \left[ A_\n(p^{(2)}), A_\m(p^{(3)})\right] \Bigr)\nn\\
&=& g_{YM}  \int d^d x ~ i\,\text{tr} \left(\p_\m A_\n - \p_\n A_\m \right) \left[ A^\m, A^\n \right] \label{gauge3}. 
\eeq 

\section{Four-Gauge Field Vertex from the Deformed Four-String Vertex}

\begin{figure}[htbp]
   \begin {center}
    \epsfxsize=0.6\hsize

	\epsfbox{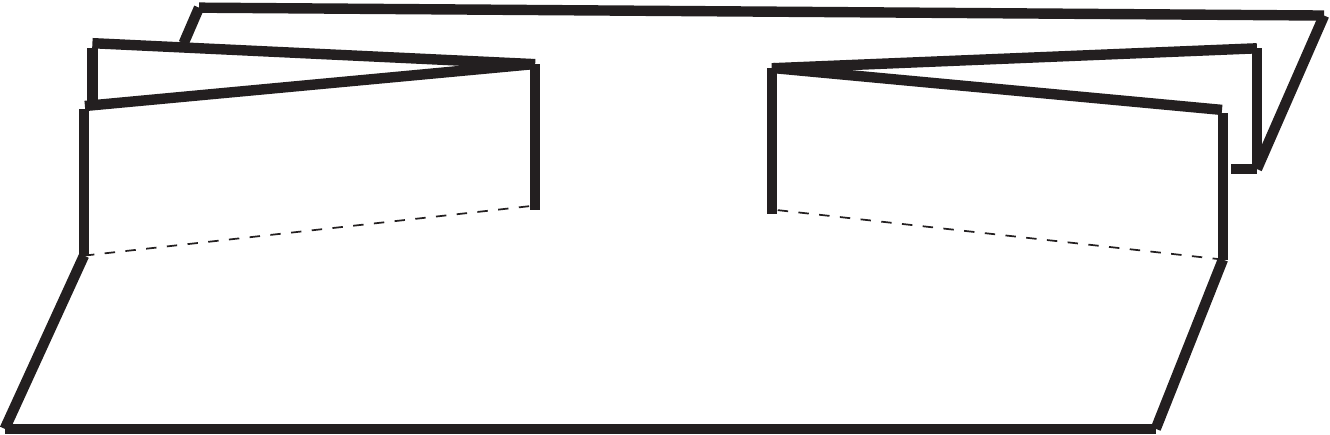}
   \end {center}
   \caption {\label{cubicfour} Four-string scattering diagram of the cubic open string field theory}
\end{figure}

\begin{figure}[htbp]
   \begin {center}
    \epsfxsize=0.6\hsize

	\epsfbox{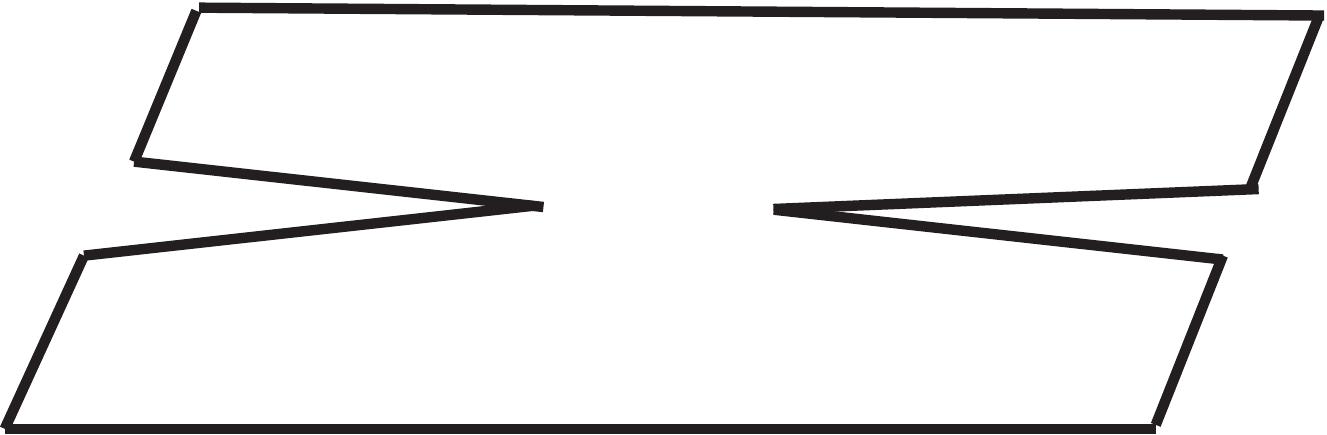}
   \end {center}
   \caption {\label{properfour} Planar diagram of the deformed four-string scattering}
\end{figure}

The four-gauge field interaction term of the Yang-Mills gauge field theory is obtained 
from the four-string scattering diagram which is perturbatively generated by the cubic interaction. 
Fig. \ref{cubicfour} depicts the effective four-string vertex of the cubic open string field theory. 
Choosing the external string states such that the physical information is encoded only on halves of 
external strings, we may effectively remove the auxiliary patches as in the case of three-string 
scattering diagram. This deformation process results in choosing the length parameters of the four strings 
as 
\beq
\a_1 = 1, ~~ \a_2 =1, ~~ \a_3 =-1, ~~ \a_4 = -1 .
\eeq
The resultant planar world sheet diagram of the deformed four-string scattering is described by 
Fig. \ref{properfour}

\begin{figure}[htbp]
   \begin {center}
    \epsfxsize=0.8\hsize

	\epsfbox{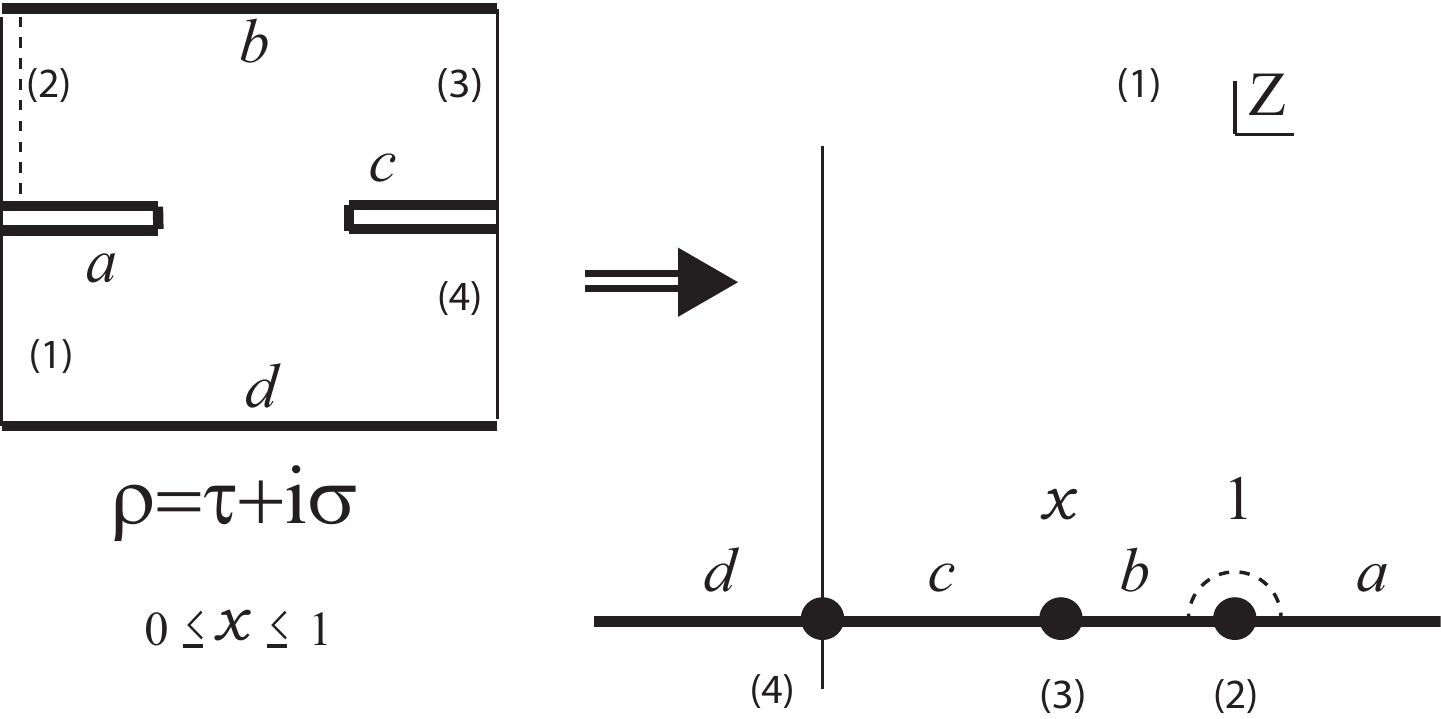}
   \end {center}
   \caption {\label{neumann4} Four-string diagram mapped onto to the upper half complex plane.}
\end{figure}

Now we shall discuss the reduction of the four-string vertex to the four-gauge field vertex in the 
zero-slope limit. 
The Witten's cubic open string field theory action does not contain a four-string interaction 
term in contrast to the light-cone string field theory and the covariantized light-cone 
string field theory of HIKKO \cite{Hata1986}. Thus, the four-gauge field interaction term of the Yang-Mills 
gauge field theory should be derived solely from the effective four-string interaction, perturbatively 
generated by the three-string interaction. Having deformed the four-string world sheet diagram into the 
planar diagram, we may map it onto the upper half complex plane as shown in Fig. \ref{neumann4} by
the following Schwarz-Christoffel transformation
\beq
\rho = \sum_{r=1}^4 \a_r \ln (z-Z_r)= \ln(1-z) - \ln z - \ln (z-x)
\eeq 
with $Z_1 = \infty, ~ Z_2 =1, ~ Z_3 = x, ~ Z_4=0$. The parameter $x$ is identified as the Koba-Nielsen variable of the four-string scattering.

We may derive the $SL(2,R)$ invariant momentum dependent four-string scattering amplitude from the three-string interaction by using the Cremmer-Gervais identity \cite{Cremmer75} as follows
\begin{subequations} 
\beq
{\cal F}_{[4]} &=& 3^2 \times \frac{1}{2!} \left(\frac{2g}{3}\right)^2 \int^{1}_0\left\vert\frac{\prod_{r=1}^4 dZ_r }{ dV_{abc}}\right\vert 
\prod_{r<s} \vert Z_r - Z_s \vert^{p_r \cdot p_s} \exp\left[-\sum_{r=1}^4 \bar N^{[4]rr}_{00} \right]\nn\\
&& ~~~~~~~~~~~~~~~~~~~~~~~\text{tr} \bigl\langle \Psi_1, \Psi_2, \Psi_3, \Psi_4 \bigl\vert \exp \left[E_{[4]} \right] \bigr\vert 0 \bigr\rangle ,\\
\left[E_{[4]} \right] \bigr\vert 0 \bigr\rangle &=&  \sum_{r, s =1}^4 \left\{ \half \sum_{m, n \ge 0} 
\bar N^{[4] rs}_{mn}\a^{(r)\dag}_{m} \cdot \a^{(s)\dag}_{n} \right\}\bigr\vert 0 \bigr\rangle . 
\eeq
\end{subequations}
If we choose the external four-string state as 
\beq
\bigl\langle \Psi_1, \Psi_2, \Psi_3, \Psi_4 \bigl\vert =\bigl\langle 0 \vert \left(\prod_{i=1}^4 A(p^{(i)}) \cdot a^{(i)}_1 \right)
\eeq
we may find that the four-string scattering amplitude yields in the zero-slope limit to following 
effective four-gauge field action:
\beq
S_{[4]} 
&=& \frac{1}{2!} 2 g_{YM}^2 \times \frac{1}{2!} \frac{1}{2^2} \times 2^3 \int \prod_{r=1}^4 dp^{(r)} \d \left(\sum_{r=1}^4 p^{(r)} \right) \int \left\vert\frac{\prod_{r=1}^4 dZ_r }{ dV_{abc}}\right\vert \prod_{r<s} \vert Z_r - Z_s \vert^{p_r \cdot p_s} \nn\\
&&
\exp\left[-\sum_{r=1}^4 \bar N^{[4]rr}_{00} \right] 
\text{tr}
\Biggl (\bar N^{12}_{11} \bar N^{34}_{11} A^\m(p_1) A_\m(p_2) A^\n(p_3) A_\n(p_4) \nn\\
&& + \bar N^{13}_{11} \bar N^{24}_{11} A^\m(p_1) A^\n(p_2) A_\m(p_3)  A_\n(p_4)
+ \bar N^{14}_{11} \bar N^{23}_{11} A^\m(p_1) A^\n(p_2) A_\n(p_3) A_\m(p_4) 
\Biggr) .
\eeq
Using $\left\vert\frac{\prod_{r=1}^4 dZ_r }{ dV_{abc}}\right\vert = Z_1^2 dx$, 
\begin{subequations}
\beq
\exp\left[-\sum_{r=1}^4 \bar N^{[4]rr}_{00} \right] \bar N^{12}_{11} \bar N^{34}_{11} 
&=& \frac{1}{Z_1^2} \frac{1}{x^2}, \\
\exp\left[-\sum_{r=1}^4 \bar N^{[4]rr}_{00} \right] \bar N^{13}_{11} \bar N^{24}_{11} 
&=& \frac{1}{Z_1^2}, \\
\exp\left[-\sum_{r=1}^4 \bar N^{[4]rr}_{00} \right] \bar N^{14}_{11} \bar N^{23}_{11} 
&=& \frac{1}{Z_1^2} \frac{1}{(1-x)^2} 
\eeq
\end{subequations}
and 
$\prod_{r<s} \vert Z_r - Z_s \vert^{p_r \cdot p_s} = x^{-\frac{s}{2}} (1-x)^{-\frac{t}{2}}$ in the zero-slope
limit, we get the effective four-gauge field action as follows
\beq
S_{[4]} &=& g_{YM}^2\int \prod_{r=1}^4 dp^{(r)} \d \left(\sum_{r=1}^4 p^{(r)} \right) \int_0^1 dx\,\, \text{tr} \Bigl(x^{-\frac{s}{2}} (1-x)^{-\frac{t}{2}} A^\m(p_1) A^\n(p_2) A_\m(p_3)  A_\n(p_4)   \nn\\
&&  + 2 x^{-\frac{s}{2}-2} (1-x)^{-\frac{t}{2}}  A^\m(p_1) A_\m(p_2) A^\n(p_3) A_\n(p_4)\Bigr).
\eeq 
Here we define the Mandelstam variables as 
\beq
s = -(p_1+p_2)^2, ~~~ t = -(p_1+p_4)^2, ~~~ u = -(p_1+p_3)^2 . 
\eeq
In the zero-slope limit 
\beq
\int^1_0 dx x^{-\frac{s}{2}} (1-x)^{-\frac{t}{2}} = 1, ~~~~
\int^1_0 dx x^{-\frac{s}{2}-2} (1-x)^{-\frac{t}{2}} = \frac{u}{s}. 
\eeq 
The resultant effective four-gauge interaction term $S_{[4]}$ does not only contain the contact 
four-gauge field interaction but also contribution of the effective four-gauge interaction 
generated perturbatively by the three-gauge field interaction of the Yang-Mills field theory. Substracting 
the effective four-gauge field interaction of the Yang-Mills theory $S_{[4]\text{massless}}$ from $S_{[4]}$
\cite{Hata1986}, 
we get the four-gauge field contact interaction of the Yang-Mills theory 
\beq
S_{\text{Gauge[4]}} &=& S_{[4]} - S_{[4]\text{massless}} \nn\\
&=& g_{YM}^2\int \prod_{r=1}^4 dp^{(r)} \d \left(\sum_{r=1}^4 p^{(r)} \right) \, \text{tr} \, 
\Biggl(A^\m(p_1) A^\n(p_2) A_\m(p_3) A_\n(p_4) \nn\\
&&+ \frac{2u}{s} A(p_1)^\m A(p_2)_\m A(p_3)^\n A(p_4)_\n
- \left( \frac{2u}{s} + 1 \right) A^\m(p_1) A_\m(p_2) A^\n(p_3) A_\n(p_4)
\Biggr) \nn\\
&=& \frac{g^2_{YM}}{2} \int d^d x \, \text{tr}\,  [A^\m, A^\n ] [A_\m, A_\n]  \label{gauge4}.  
\eeq 
Putting together the guage field interaction terms $S_{\text{Gauge[3]}}$, Eq. (\ref{gauge3}) and 
$S_{\text{Gauge[4]}}$, Eq. (\ref{gauge4}) as well as the free field action $S_{\text{Gauge[2]}}$
which may be derived easily from the kinetic term of the string field action $\text{tr} \Psi * Q \Psi$
in the zero-slope limit, 
yields the covariant Yang-Mills field action 
\beq
S_{\text{Gauge}} = \frac{1}{2} \int d^d x \, \text{tr} \, F_{\m\n} F^{\m\n} .
\eeq

\section{Conclusions}

The Witten's cubic open string field theory possesses a number of advantages over the light-cone string field theory \cite{Mandelstam1973, Mandelstam1974, Kaku1974a, Kaku1974b, Cremmer74, Cremmer75, GreenSW} and the covariantized light-cone string field theory \cite{Hata1986}: 1. The theory is covariant and invariant under 
the BRST gauge transformation. 2. The theory does not contain any other unphysical parameter 
like the length parameters except for the string coupling $g$. 3. In contrast to two other string field theories, the Witten's open string field theory does not have a quartic interaction term besides
the cubic interaction term. However, despite those advantages, it has not been fully utilized to calculate 
particle scattering amplitudes except in a few cases. The main reason is that the world sheet diagrams 
generated by the cubic string field theory are non-planar: It is difficult to find a conformal mapping 
by which the world sheet is mapped onto simple complex planes such as the upper half plane or a circular disk
without any additional conditions or structures. One needs to impose an orbifold condition to map 
the world sheet diagrams of the three-string vertex onto a circular disk \cite{Grossjevicki87a} and has to introduce branch cuts to map the four-string vertex to the upper half plane \cite{Giddings86}. 
However, even if we found maps of the world sheets to the complex planes in the cases of the three-string and the four-string scatterings, it is difficult to extend those mappings 
systematically to evaluate general multi-string amplitudes. It is also difficult to fix the relative 
strengths of the cubic gauge field interaction term and the quartic gauge field interaction term 
because there is no 
analog of the Cremmer-Gervais identity \cite{Cremmer75} which relates the three-string scattering amplitude to the four-string scattering amplitude.

In this work, we proposed a consistent deformation of the cubic string field theory by which the world sheet 
diagrams of the multi-string scattering are effectively transformed into planar diagrams. 
Having obtained planar diagrams representing the string scattering amplitudes, we can  
adopt the light-cone field theory technique to construct the Fock space representations of multi-string 
vertices systematically. By explicit calculations, we show that the three-string amplitude and the 
four-string amplitude in the zero-slope limit yield the cubic and quartic gauge interaction terms of the 
Yang-Mills theory if the external string states are chosen to be the massless gauge particles. 
The deformation process is applicable to multi-string scattering with an arbitrary number of strings.
This work may be also regarded as a proof that the string field theory in the proper time gauge \cite{Lee88, Lee2016} is 
invariant under the BRST gauge transformation. Applications of the deformed cubic string field theory 
to various scattering processes \cite{JCLee2015,Lai2016,Huang2016a,Huang2016b} will be given elsewhere.

\vskip 1cm

\begin{acknowledgments}
This work was supported by Kangwon National University. The author benefited from discussions with
Soo-Jong Rey, Yi Yang, Jen-Chi Lee, Yuji Okawa, Yu-tin Huang and participants of IBS string workshop 2016. 
Part of this work was done during author's visit to IBS (Korea) and NCTU (Taiwan). 
\end{acknowledgments}


\end{document}